\documentclass[sigconf]{acmart}
\usepackage{amsfonts}
\usepackage{amsmath}
\usepackage{nccmath}
\usepackage{graphicx}

\usepackage{dsfont}
\usepackage{booktabs}
\usepackage{multirow}
\usepackage{tabularx}

\usepackage{cleveref}
\usepackage{hyperref}
\hypersetup{colorlinks=true,citecolor=brown}
\usepackage{url}
\usepackage{xcolor}

\acmConference[WWW '24]{Companion Proceedings of the ACM Web Conference 2024}{May 13--17, 2024}{Singapore}
\acmBooktitle{WWW '24 Companion,
  May 13--17, 2024, Singapore} 
\newtheorem{observation}{Observation}

\definecolor{mygray}{rgb}{0.75, 0.75, 0.75}

\newcommand{\node}{c} % This is typically denoted by V but lowercase is so similar to u from user... 
\newcommand{\nodeall}{\mathcal{C}}
\newcommand{\edgeall}{\mathcal{E}}
\newcommand{\emb}{\mathbf{z}}
\newcommand{\embtt}{\mathbf{h}}

\newcommand{\ith}[1]{^{(#1)}}
\author{Marco De Nadai$^{1*}$\thanks{*These authors contributed equally to this research.}, Francesco Fabbri$^{1*}$, Paul Gigioli$^1$, Alice Wang$^{1}$, Ang Li$^1$, Fabrizio Silvestri$^{1,2}$, Laura Kim$^1$, Shawn Lin$^1$, Vladan Radosavljevic$^1$, Sandeep Ghael$^1$, David Nyhan$^1$, Hugues Bouchard$^1$, Mounia Lalmas-Roelleke$^1$, Andreas Damianou$^1$}
\def \authors{Marco De Nadai, Francesco Fabbri, Paul Gigioli, Alice Wang, Ang Li, Fabrizio Silvestri, Laura Kim, Shawn Lin, Vladan Radosavljevic, Sandeep Ghael, David Nyhan, Hugues Bouchard, Mounia Lalmas-Roelleke, Andreas Damianou}
\affiliation{%
\institution{$^1$Spotify
\country{Denmark, Spain, UK, USA}}
}
\affiliation{%
  \institution{$^2$Sapienza University of Rome
  \country{Italy}}
}

\begin{document}

\title{Personalized Audiobook Recommendations at Spotify \\Through Graph Neural Networks}


\renewcommand{\shortauthors}{De Nadai \emph{et al.}}

\begin{abstract}

In the ever-evolving digital audio landscape, Spotify, well-known for its music and talk content, has recently introduced audiobooks to its vast user base. While promising, this move presents significant challenges for personalized recommendations. Unlike music and podcasts, audiobooks, initially available for a fee, cannot be easily skimmed before purchase, posing higher stakes for the relevance of recommendations. Furthermore, introducing a new content type into an existing platform confronts extreme data sparsity, as most users are unfamiliar with this new content type. Lastly, recommending content to millions of users requires the model to react fast and be scalable. 
To address these challenges, we leverage podcast and music user preferences and introduce 2T-HGNN, a scalable recommendation system comprising Heterogeneous Graph Neural Networks (HGNNs) and a Two Tower (2T) model. This novel approach uncovers nuanced item relationships while ensuring low latency and complexity. We decouple users from the HGNN graph and propose an innovative multi-link neighbor sampler. These choices, together with the 2T component, significantly reduce the complexity of the HGNN model.
Empirical evaluations involving millions of users show significant improvement in the quality of personalized recommendations, resulting in a +46\% increase in new audiobooks start rate and a +23\% boost in streaming rates. Intriguingly, our model's impact extends beyond audiobooks, benefiting established products like podcasts.

\end{abstract}




\maketitle

\section{Introduction}
\label{sec:intro}

Audiobooks trace their roots in the ancient tradition of narrative: oral storytelling. Despite representing just 7\% of the broader book market, their annual consumption growth rate of 20\%~\cite{explosionaudiobooks} highlights the increasing need for personalized recommendations. Spotify, a leading audio streaming platform serving hundreds of millions of users, recently added audiobooks to its extensive catalog~\cite{explosionaudiobooks}, which already includes millions of music tracks and podcasts. While music and podcasts are consolidated on Spotify, most users are unfamiliar with the new content type. Therefore, it is challenging to develop an audiobook recommendation system that leverages scattered user interactions and seamlessly fits into the current platform.

When it comes to audiobooks, Spotify faces four main challenges. First, audiobook recommendations have not been previously studied at scale. How to best model audiobook content, understand its relationships with other audio content, and utilize available metadata for recommendations remains undetermined. Second, introducing a new content type in an existing platform faces the extreme cold-start challenge of data scarcity. 
Third, although Spotify has now included audiobooks as part of the Spotify Premium subscription\footnote{For eligible Premium users who have access to Audiobooks in selected countries~\cite{premium}.}, they were initially launched under a direct-sales model~\cite{explosionaudiobooks}. 
This sale model might influence users to have lower risk tolerance, thus creating higher stakes for the relevancy and accuracy of audiobook recommendations.
Furthermore, this model limits the volume of explicit positive interaction signals, such as streams and purchases, requiring the use of implicit signals to overcome interaction sparsity.
Finally, integrating a new product into an existing platform requires the recommendation system to be efficient, scalable, and modular. The model has to serve hundreds of millions of users with minimal latency and be flexible enough to accommodate evolving user interactions and product features. Modularity is also crucial to ensure the model's components can be adapted and reused in various projects and contexts (e.g., personalized recommendations on the home page and search).

In response to these challenges, we present 2T-HGNN, a scalable and modular graph-based recommendation system that combines a Heterogeneous Graph Neural Network (HGNN)~\cite{chen2021graph} with a Two tower (2T) model~\cite{yi2019sampling}, ensuring effective recommendations for all users with only minimal latency.

We conducted thorough data analysis and found that user podcast consumption is critical to understanding user audiobook preferences. Moreover, through data analysis, we confirm our intuition that implicit signals, such as ``follows'' and ``previews'' are beneficial to predicting future user purchases and streams. Thus, our 2T-HGNN leverages implicit and explicit signals from multiple content types to perform personalized recommendations. Our model combines the strengths of HGNN and 2T models. While the HGNN generates comprehensive long-range item representations based on content and user preferences, the 2T model enables scalable recommendations for all users and real-time serving with low latency during inference. Our solution decouples the recommendation task into an item-item component, via the HGNN, and a user-item component, via a 2T model.  This decoupling leads to a significantly smaller and tractable graph between items only, which we call \emph{co-listening graph}.
The co-listening graph and combination of a HGNN with a 2T reduces the HGNN’s inherent complexity of retrieving and aggregating neighboring nodes~\cite{jia2020redundancy, zeng2019graphsaint, guo2023linkless, zhang2020agl, ahmed2017inductive} and ensures scalability. The modularity of our recommendation system offers valuable flexibility. These modular components can be seamlessly integrated into existing models at Spotify. Additionally, this separation allows us to make adaptations and changes to the HGNN without direct user exposure or causing significant disruptions.

While leveraging an existing product (podcasts) to model a new product (audiobooks) provides significant benefits, there is an inherent imbalance favoring the existing content type in the user interactions. To address this issue, we introduce a balanced sampler that optimizes the HGNN training for multiple edge types by under-sampling the majority edge types. This graph sampler effectively captures representations for all content types and reduces training time by approximately 60\%.

\begin{figure}[t]
    \centering
    \includegraphics[width=0.75\columnwidth]{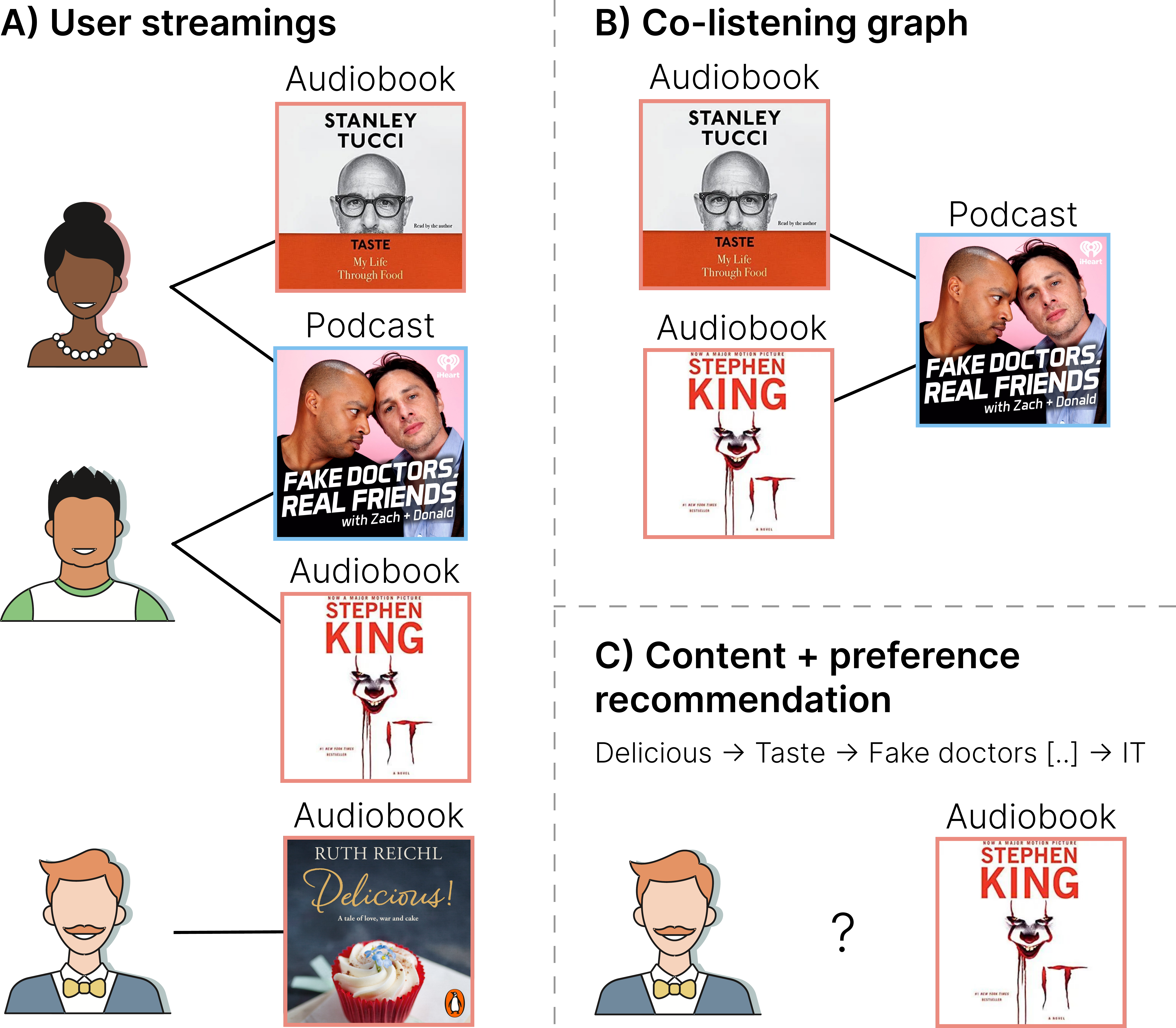}
    \caption{A) our users' consumption patterns, which involve audiobooks and podcasts; B) we build a co-listening graph with nodes representing audiobooks or podcasts, and edges connecting nodes whenever at least one user streams both; C) Audiobook \textit{IT} gets recommended because 2T-HGNN performs non-trivial recommendations using 2-hop distant patterns. \textit{Delicious} is similar to \textit{Taste}. \textit{Taste} is co-listened with \textit{Fake Doctors}, which is co-listened with \textit{IT}. } 
    \label{fig:pre-teaser}
\end{figure}

\Cref{fig:pre-teaser} overviews our model and data aggregation. Based on podcast and audiobook streaming user interactions (see \Cref{fig:pre-teaser}A), we construct the co-listening graph (see \Cref{fig:pre-teaser}B). In this graph, nodes represent audiobooks and podcasts and are connected by an edge whenever at least one user streams both. Nodes incorporate content signals from features extracted by a Large Language Model (LLM) from audiobooks and podcast descriptions. Thus, using the 2T-HGNN we build embeddings capturing non-trivial long-range dependencies, perform recommendations based on both content and user preferences (see \Cref{fig:pre-teaser}C), simultaneously learning from new (audiobooks) and more established (podcasts) content types. 

To summarize, our key contributions are: 
\begin{itemize} 
\item To our knowledge, ours is the first work to deeply investigate the design of an audiobook recommendation system at scale. 
%This content type constitutes a content type that cannot be easily skimmed before being bought, thus posing higher stakes for the relevance of recommendations.
We show how consumption of podcasts, which are usually shorter and more conversational than audiobooks, can effectively help understand user audiobook preferences. 

\item We propose a modular architecture that seamlessly integrates audiobook content into the existing recommendation system platform, combining a HGNN and 2T model in one stack. We decouple users from the graph and learn content and user preferences on a co-listening graph. The HGNN learns long-range, nuanced relations between items in the graph, while the 2T model learns user taste for audiobooks for all users, including cold-start users, in a scalable manner. 

\item To deal with the imbalance in data distribution, we first incorporate a novel edge sampler in the HGNN and then integrate the weak signals in the user representation when generating user-audiobooks predictions.

\item We conducted extensive offline experiments demonstrating the efficiency and effectiveness of 2T-HGNN. It consistently outperforms alternative methods. Furthermore, our validation using an A/B test involving millions of users resulted in a significant 23\% increase in audiobook stream rates. Remarkably, we observed a 46\% surge in the rate of people starting new audiobooks. The model is since then in production, exposed to all eligible audiobooks Spotify users.

\end{itemize}

\section{Related Work}

\textbf{Audiobooks recommendation.}
Audiobooks are part of the ``literary ecology'', along with printed books and authors~\cite{have2021reading}. Yet, they also belong to ``talk audio'' content, which includes radio and podcasts. Talk audio content is often consumed while multi-tasking such as during commuting, work, or chores~\cite{moyer2012audiobooks}. Therefore, in terms of consumption habits, audiobooks share more similarities with radio, podcasts, and even music, than with books. Nonetheless, it is currently unknown how audiobooks consumption relates to other audio content. Here, we study whether understanding podcasts consumption helps with audiobook recommendations and vice versa.\\  

\textbf{Traditional recommendation systems.}
Such systems are based on collaborative filtering approaches, which rely on capturing similarities among historical user-item interactions. These methods include matrix factorization, factorization machines, and deep neural networks~\cite{rendle2010factorization, kabiljo2015recommending, konstan1997grouplens, sarwar2001item, zhuang2013fast}. However, most collaborative approaches fall short when dealing with data sparsity. To overcome this issue, content features and additional metadata have been successful in improving recommendations.

A popular and widely adopted approach in industry, is the 2T model~\cite{yi2019sampling}. It uses separate deep neural encoders for users and items and incorporate user and item features. 2T models have found success in industrial recommendation systems, e.g.~\cite{yi2019sampling, yao2021self, yang2020mixed} and~\cite{fan2023episodes}. 
In our work we leverage a 2T architecture to guarantee scalability and fast serving performances at inference time.\\

\textbf{Graph-based recommendations.} 
Graph data structures, extensively found in online content and interaction data, provide rich information beyond traditional pairwise labels~\cite{guo2020survey}. 
Graph-based approaches have proven to be effective for recommendation task, specifically addressing challenges in cold-start scenarios and diversifying recommendations~\cite{wu2020comprehensive, chicaiza2021comprehensive}. 
For instance, DeepWalk~\cite{perozzi2014deepwalk} uses random walks to learn meaningful latent representations for social networks, while TwHIN~\cite{bordes2013translating} employs heterogeneous information networks to generate recommendations for social media. Although they are efficient in learning graph structures, these techniques are limited by their transductive nature, making them incapable of generalizing to unseen nodes~\cite{rossi2017deep, guo2020survey}.

\textbf{GNNs for recommendations.}
\label{sec:methods}
The expressive power of Graph Neural Networks (GNNs) is evident from their applications in both academic~\cite{zhang2018link, shiao2022link, velivckovic2017graph} and industrial domains~\cite{ying2018graph, sankar2021graph, gurukar2022multibisage}. 
%For instance, PinSAGE~\cite{} focuses on user-item bipartite graph recommendations at Pinterest. 
%MultiSage~\cite{} further extends this approach to multipartite graphs. 
To date, most of the current industrial GNN applications (e.g. \cite{ying2018graph, virinchi2022recommending, huang2020uber}) focus on homogeneous graphs, where nodes and edges are of a single type. Yet, in recommendation scenarios, handling diverse item types or modalities is crucial, leading to the need for Heterogeneous GNNs (HGNNs). 
However, HGNNs pose challenges as different neighbor node types have varying impacts on the node embeddings ~\cite{zhang2019heterogeneous}. 
Such imbalances require more nuanced and type-aware sampling and aggregation strategies. 

The success of (H)GNNs lies in their explicit use of neighboring (contextual) information. 
However, their large-scale adoption is limited by the complex data dependencies inherent in their neighborhood aggregation. 
To mitigate scalability and latency issues, practitioners have investigated content-only representations~\cite{ying2018graph}, graph distillation~\cite{guo2023linkless, zhang2021graph, yan2020tinygnn, xu2020graphsail}, inference speed hacks~\cite{han2015learning, zhao2020learned}, and neighborhood sampling~\cite{hamilton2017inductive}. Nevertheless, most of these methods sometimes require significant additional engineering efforts and often a compromise between accuracy and performance.

Our work presents a modular recommendation system deployed at scale at Spotify, which decouples users from HGNNs, thus requiring a leaner graph with smaller k-hop neighborhood aggregations. Our HGNN pairs with a 2T model, leveraging its proven scalability and operational speed. Moreover, we design a balanced neighborhood sampler, based on Hamilton \emph{et al.}~\cite{hamilton2017inductive} to address the imbalance between multiple edge and node types.

\section{Data \label{seq:data}}

Introducing audiobooks into Spotify, well known for music and podcasts, comes with challenges. Audiobooks were initially launched using a direct-sales strategy\footnote{Now audiobooks are available for eligible Premium subscribers who have access to Audiobooks in selected countries~\cite{premium}.}, requiring users to purchase an audiobook before it could be streamed. Thus, this severely limited the prevalence of interaction data.
Additionally, most users are unfamiliar with this new product, resulting in limited interactions and a potential bias toward more popular audiobooks.  In this section, we empirically analyze the early user interaction signals on the Spotify platform. We study the extent of our data sparsity and observe similarities between audiobooks and podcasts in terms of content or user preferences, hence motivating our approach.

\begin{figure*}[t]
    \includegraphics[width=\textwidth]{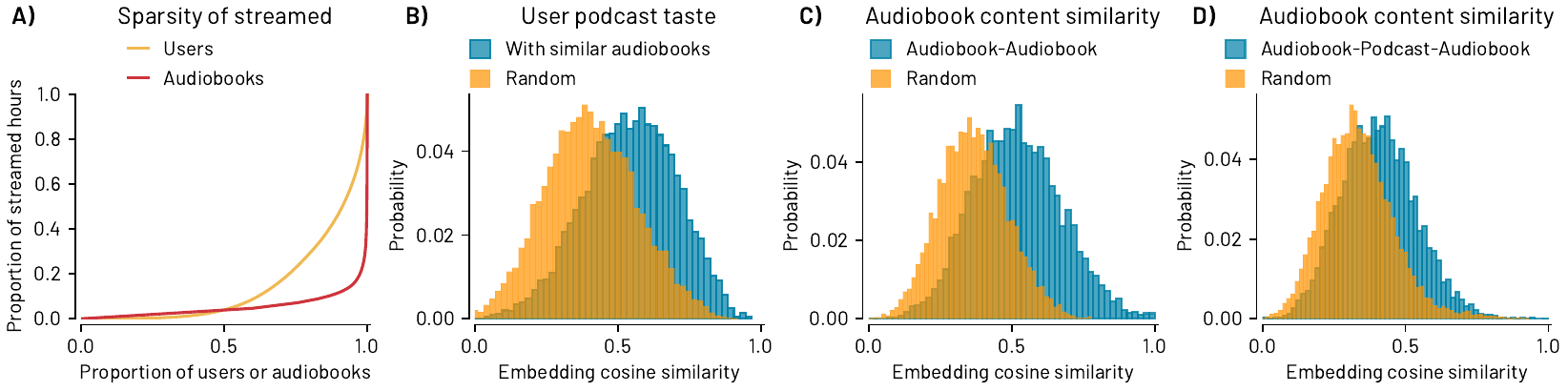}
    \caption{A) The audiobook consumption at launch is very sparse. 25\% of users account for 75\% of all streaming hours. B) Users having similar audiobook taste are more similar in podcast preferences than users selected at random.
    C) Audiobooks co-listened by at least one user have similar content embeddings (LLM embeddings extracted from the title and description of the audiobooks). D) Two audiobooks co-listened with the same podcast but not with each other have similar content embeddings.}
    \label{fig:overall}
\end{figure*}

We analyze 90 days of streaming data, comprising more than 800M+ unique streams. We focus only on podcasts and audiobooks to reduce the complexity of our analysis, since early results showed that audiobook consumption exhibits more similarity with podcast consumption than with music consumption. \Cref{fig:overall}A shows the distribution of streamed hours among users and audiobook titles. Notably, approximately 25\% of users account for 75\% of all streaming hours, and the graph illustrates that the top 20\% of audiobooks contribute to over 80\% of all streamed hours.

{\observation{Audiobook streams are mostly dominated by power users and popular titles.}}
\\

Early empirical assessments show that over 70\% of initial audiobook consumers had previously engaged with podcasts. Consequently, user interactions with podcasts could offer valuable insights into understanding audiobook user preferences. We use the Spotify podcast model currently in production to extract user embeddings, which reflect individual podcast preferences. From them, we determine whether users sharing at least one streamed audiobook exhibit greater similarity than users that streamed different audiobooks. 
To investigate this, we randomly sample 10,000 pairs $(u, u')$ of user representations in which $u$ streamed at least one audiobook that $u'$ also streamed. Then, we also randomly sample 10,000 pairs $(u'', u''')$ of user representations coupled together at random. As shown in Figure \ref{fig:overall}B, the cosine similarity between users with shared audiobook co-listenings exhibit a significantly higher level of similarity than those users coupled at random.

Content information can also provide hints about user consumption. For each audiobook in the catalog, we use text metadata (i.e., title and description) to generate low-dimensional representations  via multi-language Sentence-BERT~\cite{reimers-2019-sentence-bert}. Then, we select 10,000 distinct pairs of audiobooks in which, for each pair, at least one user listened to both audiobooks and 10,000 pairs in which audiobooks are randomly paired. Figure~\ref{fig:overall}C shows that co-listened audiobook pairs present a higher level of similarity than those that are randomly coupled, highlighting the importance of considering content metadata in the recommendation architecture.

{\observation{Podcasts user tastes and content information are informative for inferring users' audiobook consumption patterns.}}
\\

Podcast interactions help capture user taste in audiobooks, and co-listened audiobooks have higher similarity than non-co-listened ones. Thus, can podcast co-listenings serve as a reliable indicator of audiobook similarity? To answer this question, we build a co-listening graph with audiobooks and podcast nodes connected whenever at least one user co-listens them. Then, we randomly sample 10,000 pairs of audiobooks that are connected only through shared podcast co-listenings. \Cref{fig:overall}D shows that indeed sampled audiobooks connected through shared podcasts exhibit a notably stronger similarity.

{\observation{Accounting for podcast interactions with audiobooks is essential for better understanding user preferences.}}\\

Audiobook interactions are very sparse. This sparsity can be attributed to two main factors. First, most users are unfamiliar with the new content type. Secondly, users encounter a paywall when attempting to access the content, thus providing a higher barrier to stream. This also increases the imbalance of consumption signals between content types, since podcasts are freely accessible to users.

Users interact with audiobooks on the platform mainly from the home and search pages. Once a user selects an audiobook of interest, they visit the webpage and possibly follow (the updates), preview (i.e. playing a 30s sample), or show intent to pay (i.e., a purchase interaction without a completed purchase process). We refer to these collected signals as \emph{weak signals}. 

Here we investigate whether these interactions could inform future audiobook purchases and consumption. We analyze more than 198 million interactions and predict future user streams from past weak signals. We use multiple logistic regressions, one for each type of signal. %By examining the 'odds ratio' of the logistic regression, we quantify how a one-unit change in weak signals affects the likelihood of initiating a new stream. 
Results indicate that a higher occurrence of ``follow'' signals significantly boosts the odds of initiating a new stream ($+118\%$), whereas ``intent to pay'' ($+13\%$) and ``preview'' ($+18\%$) signals are also positively associated with stream initiation. We refer the reader to Appendix A for more detailed results on weak signals.

{\observation{Incorporating weak signals into our model can predict future streams and uncover subtle user preferences and intents.}}

\begin{figure*}[h!]  
\begin{center}  
\includegraphics[width=0.85\textwidth]{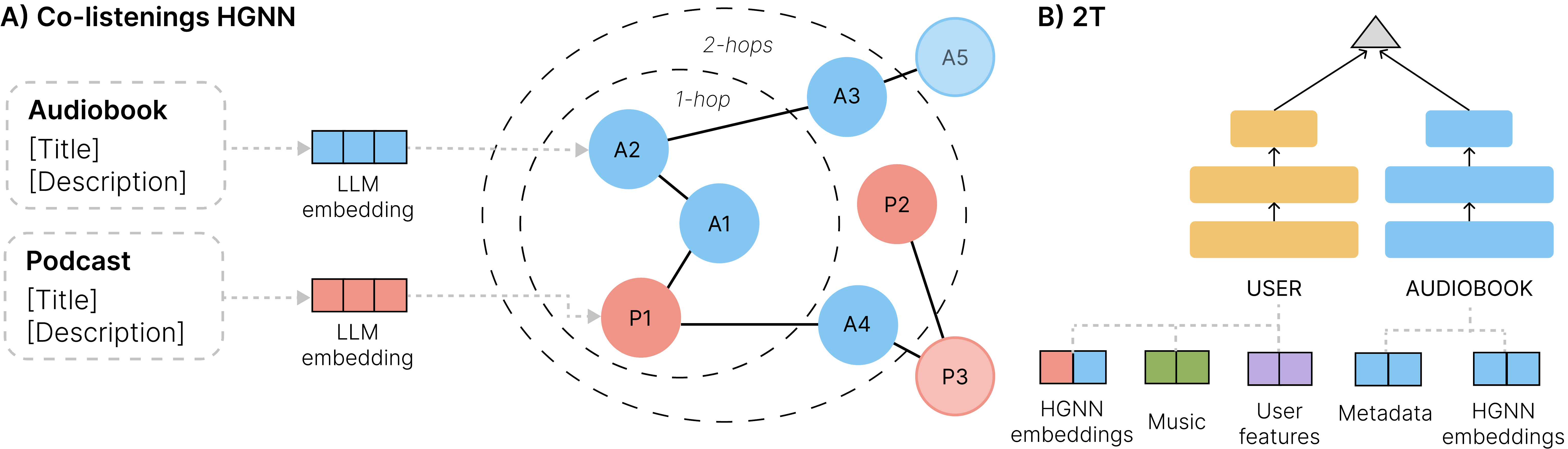}  
\end{center}  
\caption{Overview of our model. A) We represent audiobook-podcast relationships using a heterogeneous graph comprising two node types: audiobook and podcast, connected to each other whenever at least one user has listened to both. Each node has LLM embedding features extracted from the titles and descriptions of audiobooks and podcasts. We use a 2-layers HGNN on top of this graph. B) Our 2T model recommends audiobooks to users by exploiting HGNN embeddings, user demographic features (e.g. country and age), and historical user interactions (music, podcasts and audiobooks) represented as embeddings. } \label{fig:architecture_full_stack} 
\end{figure*}  

\section{Model} 
We introduce 2T-HGNN, a modular and efficient architecture for audiobook recommendations. It is modular in nature, consisting of both an HGNN and a 2T model. This modularity ensures that 2T-HGNN meets Spotify's technical requirements as outlined in~\Cref{sec:intro}, including high performance, efficiency, and flexibility in generating embeddings suitable for models deployed in various contexts such as home and search pages.

2T-HGNN addresses the audiobook interactions sparsity with a HGNN model, which is well-suited for capturing higher-order item relationships in sparse data. Our model is built upon a co-listening graph that connects content types whenever a user streams both. This graph includes both podcast and content information and incorporates co-listening interactions between podcasts as well as between podcasts and audiobooks.

The 2T builds on the audiobook and podcast representations generated by the HGNN to serve recommendations to millions of users. The HGNN and 2T can be seen as item-centric and user-centric components, respectively, working together to achieve user taste representation learning at scale. Additionally, the 2T leverages weak signals to further account for sparsity of explicit interactions (audiobook streams), thereby improving the quality of recommendations.
We refer to~\Cref{fig:architecture_full_stack} for the visual description of 2T-HGNN.

\subsection{Heterogeneous Graph Neural Network} 
\label{sec:HGNN}
HGNNs enable a comprehensive understanding of multiple data entities and relationships represented on a graph. Nevertheless, there are multiple ways to represent content and user preferences within a graph. Our approach employs a co-listening graph for content and user preferences, where users are not explicitly treated as nodes. This decoupling helps circumvent the challenges associated with HGNN neighborhood aggregations~\cite{hamilton2017inductive}, potentially involving a vast user base.
This approach guarantees the scalability and efficiency of our platform, enabling us to learn content representations from millions of items and user interactions.

\subsubsection{Graph construction} We build a co-listening graph where catalogue items $\node \in \nodeall$ (i.e. audiobooks and podcasts) constitute nodes. An edge $(\node\ith{i}, \node\ith{j}) \in \edgeall$ between two items is included if there is at least one user who interacted with both items $\node\ith{i}$  and $\node\ith{j}$. In our heterogeneous graph, each node is associated with a specific node type $s \in \mathcal{S} = \{a, p \}$, i.e. audiobook and podcast types accordingly. Further, we define a function $\phi: \nodeall \rightarrow \mathcal{S}$ mapping nodes to node types and $\langle \phi(\node), \phi(\node') \rangle$ mapping the different relationship of an edge $\epsilon = (\node, \node')$ connecting nodes $\node$ and $\node'$. Following the results in \Cref{seq:data} (Observation 2, 3), we only consider relations of the type $r \in \mathcal{R} = \{ (a, a), (a, p), (p, p) \}$, i.e. audiobook-audiobook, audiobook-podcast and podcast-podcast connections. 
By including two content types and different types of relations, we aim to capture latent connections between podcasts and audiobooks even while user interactions with audiobooks are sparse. 

To enhance our understanding of the catalog content, we incorporate node features via LLM embeddings. We use titles and description of all podcasts and audiobooks in our catalog and the multi-language Sentence-BERT model~\cite{reimers-2019-sentence-bert} to create these embeddings (see \Cref{fig:architecture_full_stack}A), which can be seen as low-dimensional representations of the content of audiobooks and podcasts. The HGNN learns complex patterns within our catalog's items from this graph, which contains information on both content and user preferences.

\subsubsection{Heterogeneous GNN design \& training:} The HGNN model is based on the GNN message-passing paradigm~\cite{kipf2016semi, hamilton2017inductive, zhou2020graph, zhang2020deep}. The heterogeneous message passing for a node $c$ is defined as:
\begin{subequations}
\begin{align}
\embtt^k_{\mathcal{N}(\node, r)} & \leftarrow  \textsc{AGGREGATE}_r^k(\{ \embtt_{\node'}^{k-1}, \forall \node' \in \mathcal{N}(\node, r) \} ) \label{eq:aggregate} \\
\embtt^k_\node & \leftarrow 
    \textsc{UPDATE}^k(
        \embtt^{k-1}_\node, 
        \{ 
            \embtt^k_{\mathcal{N}(\node, r)} 
        \}_{\forall r}
    ) \label{eq:update} 
\end{align}
\end{subequations}
where $k$ is the layer of a $l$-layers HGNN, UPDATE and AGGREGATE are differentiable functions based on $\node$'s neighbourhood $\mathcal{N}(\node, r)$. The neighborhood is defined as all nodes $\node'$ that are connected with the seed node $\node$ through a relation $r$, i.e. $(\node, \node') \in \edgeall$ and $\langle \phi(\node), \phi(\node') \rangle = r$. In \Cref{eq:aggregate,eq:update}, $\embtt_c^0 = x_c$ i.e. the node features. The node embedding is normalized to make the training more stable and allow efficient approximate nearest neighbor search $\emb_c = \embtt_c^l / ||\embtt_c^l||$ (see \Cref{sec:inference}). Having $l$-layered HGNNs allow them to learn from up to $l$-hop distant nodes (see \Cref{fig:architecture_full_stack}).

Specifically, our implementation is based on GraphSAGE~\cite{hamilton2017inductive}, in which the AGGREGATE and UPDATE operators are differentiable and parameterized with weight matrices $\mathbf{W}$. However, differently from the original paper, we here generalize those operators to the heterogeneous case. Specifically, we have:
\begin{align}
\textsc{AGGREGATE}^k_r 
    & = \text{max} \left(
            \{
                \sigma \left(
                    \mathbf{W}_r \embtt_{\node'}^{k-1} + \mathbf{b}
                    \right), 
                \forall \node' \in \mathcal{N}(c, r)
            \}
        \right) \label{eq:aggregate_form} \\
\textsc{UPDATE}^k_\node 
    & = \sigma \left(
            \mathbf{W}_{\node}^k \embtt_\node^{k-1}
            + \sum_r \embtt^k_{\mathcal{N}(\node,r)}
        \right), \label{eq:update_form}
\end{align}
where $\sigma$ is the non-linear activation function and the AGGREGATE operator is essentially a pooling operation across all neighbor embeddings which have been transformed through a neural network.

GraphSAGE defines $\mathcal{N}(\node, r)$ as a fixed-sized uniformly sampled neighborhood from $\{\node \in \mathcal{C}: (\node, v) \in \mathcal{E}\}$ , in which the sampled neighborhood is composed by different uniform samples at each training iteration. This sampling ensures that the memory and expected runtime of a single batch is limited by user-defined hyperparameters (i.e. the number of sampled nodes)~\cite{hamilton2017inductive}.

In the HGNN, the message passing and the back-propagation steps are repeated for multiple epochs, such that all parameters can be adjusted according to the training loss. In particular, we optimize the HGNN through a contrastive loss that maximizes the inner product between the anchor and a positive sample (i.e. connected nodes in the graph), while minimizing the inner product between the anchor and the negative samples. Here, the negative samples are composed by the nodes that are not connected to the anchor by an edge. We traverse all the edges of the graph, each time selecting a pair $(\emb_a, \emb_p)$ of connected nodes HGNN embeddings and randomly sample negatives $\{\emb_n | n \sim \mathcal{C}\}$ embeddings, minimizing:
\begin{equation}
    \mathcal{L}_{HGNN}(z_a, z_p) = \mathbb{E}_{n \sim \mathcal{C}} \max \{ 0, \emb_a \cdot \emb_n - \emb_a \cdot \emb_p + \Delta \}
    \label{eq:loss}
\end{equation}%
where $\Delta$ denotes the margin hyper-parameter. All nodes are sampled along with their $l$-hop sampled neighbors (Hamilton \emph{et al.}~\cite{hamilton2017inductive}).

\subsubsection{Balanced multi-link neighbourhood sampler.} 

Our co-listening graph exhibits a significant imbalance, characterized by an abundance of podcast-podcast and audiobook-podcast edges compared to audiobook-audiobook connections. Failing to consider this imbalance in our optimization process could lead our HGNN to drift away from its main task i.e. creating high quality audiobook embeddings. 

To address this imbalance, we have designed a \textit{multi-link neighborhood sampler} that  bring balance to the number of edge types minimized by \Cref{eq:loss}. It does so by reducing the number of majority edge types contained in the graph. For example, from the original graph containing $N$ audiobook-audiobook and $M$ audiobook-podcast edges, our multi-link neighborhood sampler selects only $N$ audiobook-audiobook connections and $N$ audiobook-podcast connections. The sampler undersamples multiple edge types at the same time and draws different uniform samples at each epoch to maximize dataset coverage during training.

This approach results in improved performance and produces more meaningful embeddings. Furthermore, this sampling strategy ensures a predictable expected runtime for each training epoch, which would be significantly extended to a worst case scenario of $O(|\mathcal{E}|)$. Specifically, in our use case, the number of co-listened podcasts would inevitably dominate the training process and convergence, with limited benefits for audiobook representations.

\subsection{Two Tower} 

2T-HGNN uses the 2T model to build user taste and new audiobook vectors from the HGNN audiobook and podcast representations. The 2T model is comprised of two feed-forward deep neural networks (towers), one for users and one for audiobooks (see \Cref{fig:architecture_full_stack}B). The user tower takes as input features user demographic information as well as the user's historical interactions with music, audiobooks and podcasts. Notably, interactions with music are represented by a vector that is pre-computed in-house by Spotify. Specifically, audiobook and podcast interactions are represented as the mean of the audiobook and podcast HGNN embeddings $\bar{\emb}_a$ and $\bar{\emb}_p$, corresponding to content the user interacted with in the last 90 days. Following Observation 4 in \Cref{seq:data}, we use both streams and weak signals, such as follows and previews.
The audiobook tower uses audiobook meta-data, such as language and genre, the LLM embedding from title and description, as well as the audiobook's HGNN embedding $\emb_a$. 

The 2T model generates two output vectors $\mathbf{o}_u$ and $\mathbf{o}_a$ for users and audiobooks respectively. Then, it minimizes the following loss, encouraging user vectors to be close to the audiobooks vectors they have listened to, and far away from other audiobook samples:
\begin{equation}
    \mathcal{L}_{2T}(\mathbf{o}_a, \mathbf{o}_u) = \mathbb{E}_{n \sim \mathcal{B}} \left [ \mathbf{o}_u \cdot \mathbf{o}_n - \mathbf{o}_u \cdot \mathbf{o}_a 
 \right ],
    \label{eq:loss2T} 
\end{equation}
where $\mathcal{B}$ are the in-batch negative audiobook samples. We weight the loss by the inverse probability of occurrence of items in the training dataset to prevent over-sampling popular negatives.

\subsection{2T-HGNN Recommendations} 
\label{sec:inference}

2T-HGNN generates daily user and audiobook vectors, where the audiobook vectors $\mathbf{o}_u$ are close in dot product distance to users that they will be recommended to. Each day, we first train the HGNN model and pass the resulting podcast and audiobook embeddings to the 2T model for training.  Once the 2T model is trained, we generate vectors for our audiobooks in the catalog and build a Nearest Neighbor (NN) index for online serving. Since the number of audiobooks used is relatively small, we use brute-force search to retrieve candidates from the index. As soon as the catalogue increases, we will use an approximate k-NN index~\cite{annoy} to query candidates more efficiently.  At serving time, we generate user vectors in real-time by passing user features to our user tower and querying our k-NN index to retrieve $k$ audiobook candidates for recommendation. Note that this does not preclude us to update user embeddings in real-time. Item vectors are pre-built and inserted into the index whereas user vectors are generated in real-time to be highly reactive for new coldstart users. Latency is ensured to be smaller than 100 ms.

Note that our HGNN can perform inductive inference~\cite{hamilton2017inductive}, meaning that it can generate embeddings for audiobooks that do not appear in the training co-listening graph. For example, the embedding for an audiobook that has never been streamed can be generated with just the LLM features. Moreover, the modularity of 2T-HGNN allows us to train the HGNN at a difference cadence from the 2T model training. For example, one might train the HGNN once a week to save on training costs but train the 2T model everyday to keep the user representations fresh. We leave this exploration and its impact on the performance to future investigations.

\subsubsection{Implementation details} The HGNN models have two layers and are based on GraphSAGE~\cite{hamilton2017inductive}. They are implemented in PyTorch and optimized using Adam~\cite{kingma2014adam}. We train all models with a batch size 256 and learning rate of 0.001 on a single NVIDIA T4 GPU with PyTorch Geometric~\cite{Fey/Lenssen/2019}. Training included a maximum of 50 epochs with early stopping criteria. We saved the best-performing model based on the validation set and stopped training after 10 successive epochs without improvement.

The 2T model, implemented in Tensorflow, utilized a batch size of 128 and a learning rate of 0.001 with Adam~\cite{kingma2014adam}. Each tower consists of three fully connected layers with sizes of 512, 256, and 128. Training took place on a single machine with an Intel 16 vCPU and 128 GB memory. The model was trained for 10 epochs. Other than GNN embeddings, the user tower uses demographic features (age and country) as well as interaction features (audiobook, podcast, artist) that are represented as lists of embeddings. The audiobook tower uses metadata features (i.e. language and BISAC genre code) and LLM embeddings of the title and description from Sentence-BERT~\cite{reimers-2019-sentence-bert}.  The output of each tower is a 128-dimensional vector.   

%%%%%%%%%%%%%%%%%%%%%%%%%%%%%%%%%%%%%%%

\section{Experiments and Results}

We evaluate our model performance using both offline metrics and an online A/B test, in which audiobook recommendations are exposed to real users of our platform.

\subsection{Offline Evaluation Setup}

\subsubsection{Data} For the offline evaluation, we use a large scale dataset built by collecting user interactions with podcasts and audiobooks from the last 90 days. The dataset comprises a subset of 10M users, 3.5M+ podcasts, and 250K+ audiobooks. The evaluation is done on a hold-out dataset comprising all the audiobook and podcast streams of users in the last 14 days. Thus, we split data following the gold-standard~\cite{shapira2022recommender} of a global timelime train/hold-out split scheme, in which users actions are split with a single time point split, with a time window of 14 days. The train split data was further divided in HGNN-train and HGNN-validation sets, which comprises 10\% of the train split. The HGNN training included a maximum of 50 epochs with early stopping criteria. We saved the best-performing model based on the validation set and stopped training after 10 successive epochs without improvement. 

\subsubsection{Evaluation metrics} We evaluate the performance of our recommendation task through three standard metrics namely Hit-Rate@K (HR@K), in which $K = 10$, Mean Reciprocal Rank (MRR) and catalog Coverage. We refer to \Cref{sec:metrics} for additional details.

\subsubsection{Baselines}
We evaluate our proposal on audiobook recommendations, comparing it against three different baselines. First, we employ a HGNN built upon a tripartite graph composed of user, podcast and audiobook nodes. Each edge connects a user with a podcast or audiobook whenever they stream it. We refer to this model as HGNN-w-users. Next, we train a HGNN using a co-listening graph, following \Cref{sec:HGNN}. Note that this model can only recommend audiobooks to warmstart users, meaning those who have prior interactions with audiobooks. Finally, we assess the 2T model, which employs user and audiobook towers to generate recommendations. We make user item predictions through a k-NN index. We also conduct tests on two simpler baselines, Popularity~\cite{cremonesi2010performance} and LLM-KNN. The former selects the most popular items from the catalog within the last 90 days, while the latter constructs user representations by averaging the audiobooks vectors the user has interacted (streams + weak links) with in the last 90 days.

\subsection{Offline Results}

\subsubsection{Ablation}

We conduct an ablation study on our proposed 2T-HGNN model to assess the impact of its individual components. 

First, removing our balanced multi-link neighborhood sampler leads to a 6\% drop in HR@10 (see \Cref{tab:ablation}A). The increase in coverage suggests that the recommendations span more audiobooks but faces challenges recommending the most relevant content to users.

Second, we removed weak signals from the 2T-HGNN training and inference. \Cref{tab:ablation}B shows that weak links are crucial for effective audiobook recommendations. Not only does HR@10 performance significantly decrease, but the coverage also decreases, confirming our assumption in \Cref{seq:data} (Observation 4).

Then, \Cref{tab:ablation}C-D emphasizes the significance of edges types in the co-listening graph for delivering high-quality recommendations. Omitting the podcast-podcast edges results in a 6\% decline in HR@10. Notably, \Cref{tab:ablation}D reveals that eliminating audiobook-audiobook co-listening edges leads to a substantial deterioration: a 11\% reduction in HR@10 and a staggering 57\% decline in Coverage.

Finally, we show that relying only on an homogeneous graph drastically reduces the performance (\Cref{tab:ablation}E-F). Particularly, in \Cref{tab:ablation}F we train the HGNN model on an homogeneous graph composed only of podcast to podcast connections. At inference time, we use audiobook LLM features, which are in the same latent space as the podcast ones, to inductively predict all HGNN embeddings, which are then used to train the 2T-HGNN model. Doing so, we obtain marked declines: HR@10 by 16\%, MRR by 12\%, and Coverage by 52\%. These results highlight two critical aspects: i) modelling heterogeneous content is essential; and ii) the two content types, although sharing similarities, have different user preferences.

\begin{table}[t]
	\caption{Ablation study of our model.}
    \renewcommand{\tabcolsep}{1.5pt}
    %\footnotesize
    \renewcommand{\arraystretch}{1}
	\centering
	\begin{tabularx}{\columnwidth}{@{}X rrr@{}}
	\toprule
	\multirow{2}{*}{\textbf{Model}} & \multicolumn{3}{c}{\textbf{Warmstart users}} \\ \cmidrule(lr){2-4} 
	& HR@10 $\uparrow$ & MRR $\uparrow$ & Coverage $\uparrow$ \\ \midrule
        2T-HGNN & 0.353 & 0.218 & 22.3\% \\
        \midrule
        \midrule
	%2T-HGNN w/o balance \\
	A) 2T-HGNN w/o multi-edge opt. & 0.332 & 0.214 & 24.1\% \\
        B) 2T-HGNN w/o weak signals & 0.267 & 0.182 & 17\% \\ 
        C) 2T-HGNN w/o PC-PC & 0.333 & 0.210 & 22.3\% \\
        D) 2T-HGNN w/o AB-AB & 0.312 & 0.198 & 9.4\%\\
        E) 2T-GNN (AB-AB only) & 0.329 & 0.201 & 22.1\%\\
        F) 2T-GNN (PC-PC only) & 0.294 &	0.192  & 10.6\%\\
	\bottomrule
	\end{tabularx}
	\label{tab:ablation}
\end{table}

\subsubsection{Audiobook recommendation}

We compare the performance of audiobook recommendations for warmstart and coldstart users in \Cref{tab:warmstart} and \Cref{tab:coldstart}. The former are those users who streamed, previewed, showed intent to pay, or followed an audiobook, while the latter are those who never interacted with an audiobook before. 

\Cref{tab:warmstart} shows the quantitative evaluation for those users who interacted at least one time with  audiobooks. The popularity baseline performs quite well, highlighting the popularity bias issue observed in \Cref{seq:data} (Observation 1). LLM-KNN excels in coverage and MRR and shows that content-based recommendations (i.e., through similarities of audiobook descriptions) are essential in audiobook recommendations. However, this method struggles to suggest relevant (personalized) content in the first ten items (HR@10 is 0.164). In contrast, the HGNN model improves HR and MRR of 57\% and 10\% respectively over LLM-KNN, with only a marginal reduction in coverage (-3\%). This outcome suggests that HGNNs are adept at capturing subtle nuances in user preferences, which co-listening edges might effectively capture. Thus, it is essential to concurrently model both content and user preferences.

Despite outperforming LLM-KNN, HGNN-w-users exhibits sub-optimal performance in MRR and Coverage, with declines of 30\% and 53\% from the HGNN result, respectively. This decline in performance is likely attributed to the high sparsity of the user graph, characterized by a substantial number of non-connected components and a lower average degree than the co-listening graph.

Next, we compare the 2T model, which performs worse than HGNN-w-users and HGNN in all metrics. However, it requires significantly less training time and lower inference latency, positioning it as a competitive choice in the trade-off between online performance and evaluation metrics.

Thus, we finally evaluate our proposed 2T-HGNN method, which outperforms all models in HR@10, improving the best baseline by 36\%. Although its MRR and Coverage don't match the HGNN ones, it balances the recommendation performance of the HGNN model with the inference speed of the 2T-HGNN, which makes it the perfect candidate for serving millions of users in real-time recommendations. Particularly, this model improves the 2T performance by 52\%, 26\% and 5\% on HR@10, MRR and Coverage respectively.

We also evaluate 2T-HGNN improvements on long-tail recommendations by categorizing audiobooks into five popularity tiers. Tiers 3, 4, and 5, representing less popular content, are considered the long tail. The results show a significant improvement of 2T-HGNN, with HR@10 and MRR increasing by 118\% and 102\%, respectively, at no expense of Coverage.

\Cref{tab:coldstart} confirms the consistency of our findings in HR@10 and MRR for cold-start audiobook recommendations. 
This table shows the popularity bias issue worsen as the Popularity baseline surprisingly outperforms the 2T model in HR@10: the ten most popular audiobooks are often picked up by users as their primary choice for the first streamed audiobook (see \Cref{fig:overall}A). The combination of 2T+GNN continues to exhibit high performance, improving upon the 2T model by 48\% percent. However, a significant contrast emerges among the models in terms of coverage. HGNN-w-users achieves a mere 6.4\% coverage, indicating that its recommendations are limited to a small subset of the catalog. Although 2T-HGNN nearly doubles this coverage to 12.0\%, it is surpassed by the 2T model, which performs 60\% better in this regard. In other words, 2T-HGNN excels in making precise and accurate predictions, but its recommendations are limited to a narrower subset of the catalog. We do not consider this thade-off as a major issue at the moment, but something to be eventually re-consider in the future.

\subsubsection{Podcast recommendation}
Integrating the representation of audiobooks and podcasts within a single graph enables us to learn content similarities and capture user preferences across both products. Leveraging this hypothesis, we incorporated audiobooks into our existing online platform that previously featured only podcasts. Consequently, we evaluate whether the newly proposed 2T-HGNN model enhances podcast recommendations.

\Cref{tab:podcasts} reveals that the 2T-HGNN model outperforms the 2T model, the current recommendation system in production, by a margin of 7\% in HR@10 and, remarkably, it increases Coverage by 80\%  for warm and coldstart users. While the MRR performance of the model is on par with existing the model, \Cref{tab:podcasts} shows that recommendations for a pre-existing product (i.e., podcasts) can be improved by exploiting data from a distinct product (i.e., audiobooks), thereby deepening our understanding of user preferences.

\begin{table}[t]
	\caption{Audiobook recommendations for warmstart users.}
    \renewcommand{\arraystretch}{1}
     %\footnotesize
	\centering
	\begin{tabularx}{\columnwidth}{@{}Xrrrr@{}}
	\toprule
	\multirow{2}{*}{\textbf{Model}} & \multicolumn{3}{c}{\textbf{Warmstart users}} \\ \cmidrule(lr){2-4} 
	& HR@10 $\uparrow$& MRR $\uparrow$ & Coverage $\uparrow$\\ \midrule
        Popularity & 0.150 & 0.100 & 0.0\%\\ 
	LLM-KNN & 0.164 & 0.202 & 54.7\% \\
        HGNN & 0.258 & 0.224 & 52.8\% \\ 
        HGNN-w-users & 0.238 & 0.163 & 25.3\% \\
        2T & 0.231 & 0.173 & 21.2\%\\
        2T-HGNN & 0.353 & 0.218 & 22.3\% \\
	\bottomrule
	\end{tabularx}
	\label{tab:warmstart}
\end{table}

\begin{table}[t]
	\caption{Audiobook recommendations for coldstart users.}
    \renewcommand{\arraystretch}{1}
    % \footnotesize
	\centering
	\begin{tabularx}{\columnwidth}{@{}Xrrr@{}}
	\toprule
	\multirow{2}{*}{\textbf{Model}} & \multicolumn{3}{c}{\textbf{Coldstart users}}  \\ \cmidrule(lr){2-4} 
	& HR@10 $\uparrow$& MRR $\uparrow$& Coverage $\uparrow$\\ \midrule
        Popularity & 0.161 & 0.100 & 0.0\%\\ 
        HGNN-w-users & 0.174 & 0.153 & 6.4\% \\
        2T & 0.135 & 0.146 & 19.3\%\\
        2T-HGNN & 0.200 & 0.156 & 12.0\% \\
	\bottomrule
	\end{tabularx}
	\label{tab:coldstart}
\end{table}

\begin{table}[t]
	\caption{Podcast recommendation performance.}
    \renewcommand{\arraystretch}{1}
     %\footnotesize
	\centering
	\begin{tabularx}{\columnwidth}{@{}Xrrr rrr@{}}
	\toprule
	\textbf{Model} & \textbf{HR@10 $\uparrow$} & \textbf{MRR $\uparrow$} & \textbf{Coverage $\uparrow$} \\ \midrule
        Popularity & 0.059 & 0.100 & 0.0\%\\ 
        2T & 0.114 & 0.135 & 11.4\%  \\
        2T-HGNN & 0.123 & 0.138 & 20.6\%\\
	\bottomrule
	\end{tabularx}
	\label{tab:podcasts}
\end{table}

\subsection{Production A/B Experiment}

We run an A/B experiment using 2T-HGNN as a candidate generator to better understand the online performance of the model. The focus of the experiment is ``Audiobook for you'', a section of the Spotify home page that shows the top $k$ audiobooks personalized recommendations. 
This experiment involved a sample of 11.5 million monthly active users, who were randomly divided into three groups. The first one was exposed to the model currently in production, the second group received recommendations generated by a 2T model, while the third one from the 2T-HGNN model. We tested the 2T model as a competitive alternative to the 2T-HGNN. All models are trained on the same date range of data for fair comparisons.

\Cref{tab:abtest} shows that 2T-HGNN significantly increased new audiobook start rate and led to a higher audiobook stream rate. In contrast, the 2T model had a lower uplift in audiobook start rate and did not produce a statistically significant change in stream rate.

\begin{table}[t]
	\caption{Online A/B test results. }
    \renewcommand{\arraystretch}{1}
     %\footnotesize
	\centering
	\begin{tabularx}{\columnwidth}{@{}Xrr@{}}
	\toprule
	\textbf{Model} & \multicolumn{2}{c}{\textbf{Business metric}} \\ \cmidrule(lr){2-3} 
	& Stream rate & New audiobooks start rate \\ \midrule
        2T & Neutral & +23.87\%  \\
        2T-HGNN & +25.82\% & +46.83\% \\ 
	\bottomrule
	\end{tabularx}
	\label{tab:abtest}
\end{table}

\section{Conclusions}
In this work we introduce the architecture powering personalization of audiobook recommendations in Spotify. We propose 2T-HGNN, a model that effectively captures users' taste for audiobooks through the combination of a HGNN architecture and a 2T model. Our modular approach allows us to decouple complex item-item relationships (through the HGNN) while producing scalable recommendations for all users (through the 2T). 
Our results reveal a strong connection between user preferences for audiobooks and podcasts. Notably, modelling the two content types together improve the recommendation quality of both content types. Our online A/B test demonstrates the success of deploying 2T-HGNN for audiobook recommendations and, more generally, its ability to power recommendations for a new talk audio product on an existing platform. The model is now in production and exposed to millions of users. We believe this approach can scale across various content types leading to a better personalized experience for online users.

\section{Acknowledgments}
F.S. thanks all these projects for partially supporting this work: FAIR (PE0000013) and SERICS (PE00000014) under the MUR National Recovery and Resilience Plan funded by the European Union - NextGenerationEU, the ERC Advanced Grant 788893 AMDROMA, EC H2020RIA project “SoBigData++” (871042), PNRR MUR project IR0000013-SoBigData.it and project NEREO (Neural Reasoning over Open Data) project funded by the Italian Ministry of Education and Research (PRIN) Grant no. 2022AEFHAZ.

\newpage

%\clearpage
\bibliographystyle{abbrv}
\bibliography{main}

\appendix
\section{Evaluation metrics}
\label{sec:metrics}
We evaluate the performance of our recommendation task on implicit feedback through two standard metrics namely HR@K and MRR. The former measures the proportion of users for whom at least one relevant item (the one chosen by the user) has been recommended in the top $K = 10$ items (see \Cref{eq:HR})), while the latter takes into account how far the item the user interacted is in the list of recommended items (see \Cref{eq:MRR}). We also evaluate the catalogue coverage of our recommendations, which helps understand the long-tail recommendation issue and whether the recommendation system can ameliorate popularity bias (see \Cref{eq:coverage}). 

\begin{equation}
\text{HR}@K = \frac{\sum_{u \in U}{\mathds{1}(\text{the relevant item is in top K})}}{|U|}
\label{eq:HR}
\end{equation}
\begin{equation}
    MRR = \frac{1}{|U|} \sum_{u \in U} \frac{1}{r_u}
\label{eq:MRR}
\end{equation}
\begin{equation}
    Coverage = \frac{|\cup_{u \in U} \Upsilon_u|}{|\Upsilon|}
\label{eq:coverage}
\end{equation}
where $U$ is the set of users $r_u$ is the rank of the relevant item, $\Upsilon_u$ is the set of items recommended to user $u$, and $\Upsilon$ is the entire catalogue. For performance reasons, we limit the set of recommended items to the first 100 recommended items for MRR and Coverage.

\section{Weak signals co-occurences}

We here explore the concept of \emph{weak signals}, which refer to user actions performed prior to completing an audiobook purchases. We focus on three specific actions: "follow", which allows users to keep up with updates of an audiobook; "preview", enabling users to listen to a 30-second sample of the audiobook; and "intent to pay", signaling an incomplete purchase attempt. Our aim is to assess the informativeness of these weak signals by analyzing over 198 million interactions, examining their co-occurrences and predictive value concerning a user's initial streaming activity.

\begin{figure}[h!]
\begin{tabular}{c}
\centering
             \includegraphics[width=1.0\columnwidth]{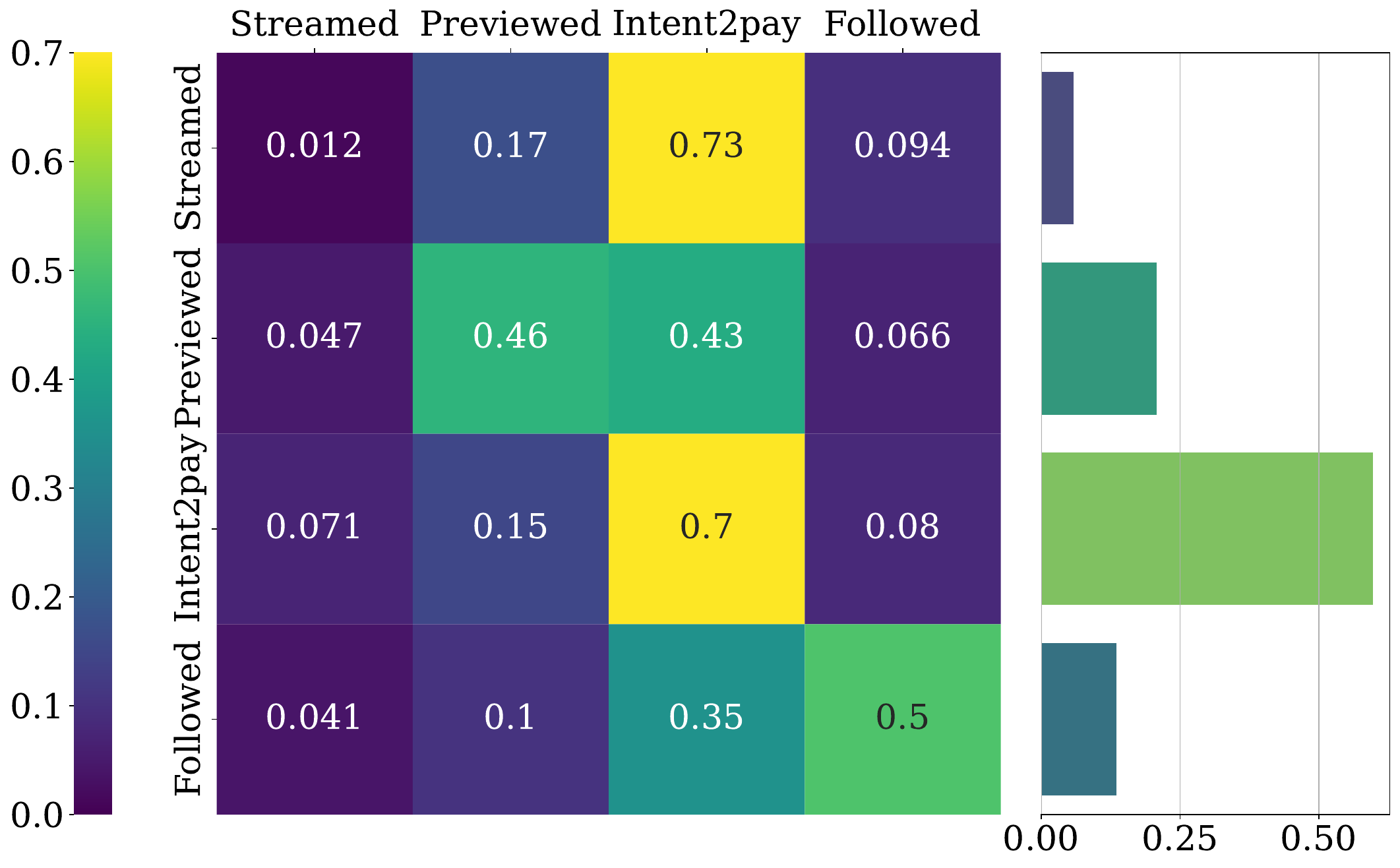}
\end{tabular}
\caption{Co-occurrence Patterns among weak signals. The heatmap illustrates the distribution of signal co-occurrences, with each $(i,j)$ entry representing the fraction of occurrences of signal $j$ in relation to the total occurrences of signal $i$. The adjacent bar plot on the right provides insights into the relative distribution of signals within rows.}
\label{fig:heatmap_stats}
\end{figure}

\Cref{fig:heatmap_stats}, how these signals co-occur, with each row representing the distribution of a signal in conjunction with those in the columns.
Each row of the barplot highlights the proportion of interactions involving that particular signal, offering insight into its relative significance within the total dataset.

The findings indicate that interactions signaling "intent to pay" are strongly linked with the primary stream, frequently occurring in conjunction with a purchase. 
Although "follow" interactions are less common, they do not often coincide with other signals. Similarly, "preview" interactions, despite their infrequency, demonstrate a moderate rate of co-occurrence with other types of interactions. This analysis sheds light on the potential of weak signals as indicators of user engagement and purchasing behavior.

\end{document}